\documentclass[amssymb,prb,twocolumn,showpacs,superscriptaddress]{revtex4}
\usepackage{epsfig}
\usepackage{dcolumn}
\usepackage{amsmath}
\begin{document}

\title{Interaction induced magnetic field asymmetry of nonlinear 
mesoscopic electrical transport} 
\author{Markus B\"uttiker}
\affiliation{D\'epartement de Physique Th\'eorique,
Universit\'e de Gen\`eve, CH-1211 Gen\`eve 4, Switzerland}
\author{David S\'anchez}
\affiliation{Departament de F\'{\i}sica,
Universitat de les Illes Balears, E-07122 Palma de Mallorca, Spain}
\affiliation{D\'epartement de Physique Th\'eorique,
Universit\'e de Gen\`eve, CH-1211 Gen\`eve 4, Switzerland}
\date{\today}

\begin{abstract}
We demonstrate that the nonlinear $I$--$V$ characteristics of a two probe 
conductor is not an even function of magnetic field. While the 
conductance of a two-probe conductor is even in magnetic field, we find 
that already the contributions to the current which are second order 
in voltage, are in general not even. This implies a departure from the 
Onsager microreversibility principle already in the weakly nonlinear regime. 
Interestingly, the effect that we find is due to the Coulomb interaction.
A measurement of the magnetic field asymmetry can be used to determine the 
effective interaction strength. As a generic example, we discuss the 
$I$--$V$ characteristics of a chaotic quantum dot. The ensemble averaged $I$--$V$
of such a cavity is linear: nonlinearities are due to quantum interference.
Consequently, phase-breaking reduces the asymmetry. We support this 
statement with a calculation which treats inelastic scattering with the help
of a voltage probe. 
\end{abstract}

\pacs{73.23.-b, 73.50.Fq, 73.63.Kv}
% Keywords:
% 73.23.-b -> Electronic transport in mesoscopic systems 
% 73.50.Fq -> High-field and nonlinear effects
% 73.63.Kv -> Quantum dots
% Rectification
\maketitle

\section{Introduction}

The linear transport regime is governed by a number of 
fundamental principles. It is interesting to ask whether 
these principles extend into the non-linear regime and 
if not what causes their breakdown. For instance, the Onsager-Casimir 
theory of irreversible processes \cite{ons31,cas45} applied to 
electrical conduction \cite{cas45} 
implies that the conductance of a two probe conductor 
is an even function of magnetic field $G(B) = G(-B)$. 
We might expect that this symmetry is valid 
even in the non-linear regime and that quite generally 
the current is an even function of magnetic 
field $I(B,V) = I(-B,V)$ for any voltage $V$ applied to the
conductor. However, a closer inspection shows that this is not true.
In a recent work~\cite{san04}, 
we have shown that already the weakly nonlinear current response quadratic 
in voltage is in general not an even function of magnetic field. 
This implies that the Onsager relations are 
{\it strictly} limited 
to the linear transport regime and that the $I$--$V$
characteristics contains 
a term \cite{spi04} proportional to the magnetic field $B$ and the applied voltage squared, $BV^{2}$.
Interestingly, the effect depends on the fact that electrons 
are interacting particles. A theory based on 
free electrons gives an $I$--$V$ characteristics which 
is even in magnetic field. Consequently this discussion demonstrates 
that also the noninteracting picture of electrical transport is 
strictly limited to the linear transport regime. 

There has been a growing interest in the investigation of the nonlinear
current through low-dimensional electronic structures in which
electron transport is phase coherent.
At the same time, high-field effects have found much interest in the rapidly developing field
of molecular transport. We highlight the observation of
{\em rectification} effects in solid-state 
mesoscopic junctions~\cite{son98,lin00,sho01}
and molecular structures~\cite{mar93,che99,xue}
which, when the temperature is low enough, 
arise already at small bias voltages.
Our work\cite{san04} has been focussed
on the asymmetries shown in the rectification current of mesoscopic
systems when a magnetic field is applied in addition to the external dc bias.

\begin{figure}[b]
\centerline{
\epsfig{file=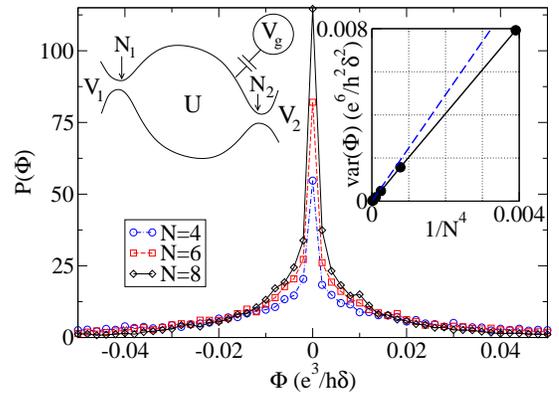,angle=0,width=0.40\textwidth,clip}
}
\caption{Magnetic field asymmetry of nonlinear two-terminal transport. A
chaotic cavity is coupled via quantum point contacts with $N_1$ and $N_2$
transverse modes to reservoirs and via a capacitance $C$ to a gate (left inset).
$P(\Phi)$ is the probability to find a cavity with a certain asymmetry $\Phi$
[see Eq. (\ref{eq_asg})] in the presence of a magnetic field generating a flux of the 
order of $h/e$. 
The variance of $\Phi$ as a function of the total channel number 
$N = N_1 + N_2$ is shown 
in the inset to the right. After Ref.~\onlinecite{san04}. 
}
\label{dot}
\end{figure}

We have first investigated a simple example of a resonant impurity 
in the quantum Hall regime.~\cite{san04} The Hall potential is {\it odd}
under magnetic field reversal. Consequently, it is sufficient 
to demonstrate that nonlinear transport even in a two probe geometry 
is sensitive to the Hall potential in order to generate an $I$--$V$ characteristics
that contains terms that are odd under magnetic field reversal. 
We have analyzed a model with a quasi-localized
state coupled to chiral edge states~\cite{for94}.
We have shown that the $I$--$V$ curve
exhibits a magnetic field asymmetry if either the charges on the edge 
states couple asymmetrically to the impurity state ({\it electrical asymmetry})
or if the impurity is coupled asymmetrically to the edge states simply due to
asymmetric tunnel couplings ({\it scattering asymmetry}).

A second system that we have examined consists of a chaotic cavity 
coupled via quantum point contacts to metallic contacts
(see left inset of Fig.~\ref{dot}). An ensemble of chaotic 
cavities, in which individual members differ only by small changes in shape, exhibits an ensemble averaged
$I$--$V$ characteristics which is linear in voltage. In contrast an individual cavity
exhibits a nonlinear $I$--$V$ characteristics due to quantum interference effects. 
An $I$--$V$ characteristics which is odd under magnetic field is a consequence of
the magnetic field asymmetry of the electrical potential landscape inside the cavity. For chaotic cavities, this effect
is the stronger the better the local potential neutralizes excess charge which 
might arise due to the applied voltage. In our work,~\cite{san04} the cavity is ballistic
(the only scattering arises due to the potential walls of the cavity and 
possibly due to diffraction of the carriers at the contacts to the cavity).    
Alternatively, one can consider a small conductor which is metallic 
diffusive.\cite{spi04} Even though a ballistic cavity and a small metallic diffusive
sample are very different conductors, the theory gives qualitatively similar predictions. Both for the ballistic cavity and the diffusive cavity the asymmetry 
becomes smaller very rapidly with increasing conductance. 
For the ballistic cavity we found detailed results in terms of the number of open channels of the quantum point contacts linking the sample to the metallic contacts. 
The effect we find is very small, at best of the order of the conductance 
quantum. Possibly, this is connected to the fact that the 
ensemble averaged $I$--$V$ characteristics of a chaotic cavity is linear. We can not exclude 
that there are special geometries that exhibit a magnetic field asymmetry which is 
much stronger than the effect reported here.
Below, we investigate the effect of inelastic scattering 
with the help of a voltage probe and demonstrate that for the chaotic cavities it reduces further 
the magnetic field asymmetry. 

The magnetic field symmetry of transport coefficients is 
an important probe of our 
understanding of transport phenomena. It is well known 
that the magnetic field symmetry of linear conductance 
played an important role in the development of a scattering theory 
that is applicable to a wide range of experimental configurations \cite{but86}.
The magnetic field symmetry of ac-transport conductances has 
similarly been discussed \cite{but93,chr} 
but only a few experimental works are available \cite{chen,som}. 
The self-consistent, nonlinear scattering approach to dc-transport which we will use here 
has been developed in Ref. \onlinecite{but93} and by 
Christen and B\"uttiker \cite{chr96}. For closely related discussions 
see Refs. \onlinecite{wang,mbds}. 
In addition to linear and nonlinear dc-transport, and ac-conductances, 
the magnetic field symmetry is also of interest for the dc-currents 
generated through pumping. 
A dynamic modulation of the shape of the 
chaotic cavity has been experimentally shown to induce a 
dc current asymmetric\cite{dic03} in $B$. The theory of magnetic field symmetry of
pumped currents is the subject of Refs.~\onlinecite{shu00,bro01,vav01,kim}. 
The magnetic field symmetry of an adiabatic two-parameter 
pump in the presence of a 
dc-voltage is discussed by Moskalets and B\"uttiker \cite{mos05}. 

The nonlinear conductance fluctuations
asymmetric in $B$ have been measured in experiments 
in chaotic cavities~\cite{zum}. 
The experiment finds fluctuations which are
smaller than predicted by the theoretical model. Since in a realistic
situation dephasing time might be of the same order as the dwell time inside
the cavity, we report below an investigation of the role of dephasing.

\section{Scattering theory of weakly nonlinear transport}

We consider a generic mesoscopic conductor connected to
$\alpha=1,\ldots , M$ reservoirs and gates. We model the transport
with the scattering approach. In this picture,
the probability amplitude for an electron to go from lead $\beta$
to lead $\alpha$ is given by the quantum-mechanical
scattering matrix $s_{\alpha\beta}(E)$, which
is a function of the carrier's energy $E$. 
It describes elastic scattering within the conductor, assuming that
inelastic processes take place only in the reservoirs far away
from the conductor. The reservoirs are maintained at thermal
equilibrium with Fermi distribution function $f(E-\mu_\alpha)$,
where $\mu_\alpha$ is the electrochemical potential.
Thus, due to microscopic
reversibility the amplitude for an electron to be transmitted from
lead $\beta$ to lead $\alpha$ at a given $B$ equals
the amplitude from lead $\alpha$ to lead $\beta$ when
the magnetic field is {\em reversed}, i.e.,
$s_{\alpha\beta}(E,B)=s_{\beta\alpha}(E,-B)$. It follows that
the two-probe linear conductance
\begin{equation}
G = \frac{e^2}{h}\int dE\, {\rm Tr}( s_{12} s_{12}^\dagger)
(-\frac{\partial f}{\partial E}) \,,
\label{eq_linearg}
\end{equation}
fulfills the Onsager's symmetry $G(B)=G(-B)$.

For linear transport the scattering matrix is evaluated 
for electron motion in the equilibrium potential
$U_{\rm eq}(\vec{r})$. 
For a nonequilibrium situation, the scattering matrix is 
a function of the potential generated by the
charges piled up in the mesoscopic conductor~\cite{but93}.
The 
potential $U(\vec{r}, \{V_\gamma\})$ now depends on
the set of electric potential shifts $\{V_\gamma\}$ applied
to the external reservoirs and nearby gates. [In what follows, 
we assume for simplicity that $U$ is uniform inside the sample.
The full theory~\cite{but93,chr96} takes into account spatial inhomogeneities
$U(\vec{r},\{V_\gamma\})$]. The electrostatic potential is essential to
a charge-conserving transport theory since the resulting
current--voltage characteristics must obey gauge invariance,
i.e., the current expressions must depend
only on voltage differences~\cite{but93,chr96}. 
Therefore, the Onsager's symmetries of the
scattering matrix $s_{\alpha\beta}(E,U)$ depend crucially
on whether or not the screening potential is an even function
of $B$. 
In linear response, Eq.~(\ref{eq_linearg}), $s_{12}$
is evaluated at the equilibrium potential
$U_{\rm eq}$, thereby the conductance is an even function
of the magnetic field, as should be. In the multiterminal case,
a generalized reciprocity theorem holds~\cite{but86}.
In contrast, in the nonlinear regime the potential is not an even 
function of magnetic field \cite{but93,chr96} as we will now discuss.

To ensure charge conservation the potential
$U$ of the conductor floats up and down in response to the
density of carriers injected through the leads. In addition,
interaction with nearby gates causes a response of $U$. As a consequence,
the electron density in the sample feels a potential
which depends on the shifts ${V_\gamma}$ of both
bias and gate voltages. To leading order in the voltage shifts
we can write
\begin{equation}
U= U_{\rm eq} + \sum_{\alpha} u_{\alpha} V_{\alpha} + {\cal O}(V^{2})\,,
\end{equation}
where the {\em characteristic potentials}~\cite{but93} $u_\gamma =
(\partial U/\partial V_{\gamma})_{\rm eq}$ relate the variation
of the screening potential in the sample
to a voltage shift in the contact $\gamma$.

To compute $u_\gamma$ we need the charge density in response to a variation 
of the voltage at the contact and need the charge response to a small variation
of $U$. These response functions can be obtained from the {\it partial density of
states},
\begin{equation}
\frac{dn_{\alpha\beta}}{dE}= -\frac{1}{4\pi i}
{\rm Tr}\left(s^{\dagger}_{\alpha\beta}
\frac{\partial s_{\alpha\beta}}{\partial eU}
- 
\frac{\partial s^{\dagger}_{\alpha\beta}}{\partial eU} s_{\alpha\beta}
\right)\,
\end{equation}
which is the portion of the density of states (DOS) associated 
with carriers that
enter the conductor from contact $\beta$ and leave it through
contact $\alpha$. (When the spatial dependence is needed, the partial
derivative with respect to the potential $U$ should be replaced by a functional
derivative). The {\em injectivity} of lead $\beta$ describes
the DOS of those carriers which are injected from lead $\alpha$
regardless to which reservoir the carriers are finally scattered,
i.e., $\overline{D}_{\beta}=\sum_\alpha dn_{\alpha\beta}/dE$. 
The {\em emissivity} of lead $\alpha$,
$\underline{D}_{\alpha}=\sum_\beta dn_{\alpha\beta}/dE$,
contains information
only about the carriers which are leaving the sample through contact 
$\alpha$ irrespective
of the injecting contact(s). Knowledge of these quantities
is essential to find the charges which pile up in the conductor
in response to a voltage shift. Since imposing charge conservation
involves a balance for the internal potential and
the scattering matrix depends itself on $U$,
this problem needs to be solved {\em self-consistently}.
We postpone the solution for the chaotic cavity to the next 
section.

An expansion of the current through lead $\alpha$
in powers of the applied voltages, 
\begin{equation}
I_\alpha=\sum_\beta G_{\alpha \beta} V_{\beta} +
\sum_{\beta\gamma} G_{\alpha\beta\gamma} V_{\beta} V_{\gamma} \,,
\end{equation}
contains a linear term given by the linear conductance conductance 
matrix $G_{\alpha \beta}$, and a leading-order rectification term
$G_{\alpha\beta\gamma}$ which reads, for spinless electrons,~\cite{chr96}
\begin{equation} 
\label{gnonl}
G_{\alpha\beta\gamma}= \frac{e^2}{h} \int dE \,
(-\frac{\partial f}{\partial E})
\frac{\partial A_{\alpha\beta}}{\partial U}
[2 u_\gamma - \delta_{\beta\gamma}]\,,
\end{equation}
where $A_{\alpha\beta}={\rm Tr}
[\mathbf{1}_{\alpha\beta}\delta_{\alpha\beta}-
s_{\alpha\beta}^\dagger s_{\alpha\beta}]$ with
${\bf 1}$ the identity matrix. Although Eq. (\ref{gnonl})
gives only the initial departure from linear behavior, the 
expansion has the advantage that all quantities (like the linear 
conductance) are evaluated 
in the equilibrium state. Therefore, from a statistical mechanical
point of view, we are on safe grounds. We now specialize to the
two-terminal case: $I\equiv I_{12}=-I_{21}$. 
The conductance matrix is then given by 
$G_{11} = G_{22}= -G_{12} = - G_{21} = G$
with $G$ given by Eq.~(\ref{eq_linearg}).
The nonlinear $I$--$V$ characteristics depends not only 
on the voltage shifts of the contacts which permit 
particle 
exchange but also on how we shift the voltage on the 
nearby gates. To be definite, we assume here $V_1 = V$
and $V_2 = V_g = 0$. Then,
to second-order in the bias voltage $V$ the
{\em differential} conductance takes the following form:
\begin{equation}
\label{eq_c2t}
\mathcal{G}\equiv\frac{dI}{dV} = G_{11} + 2 G_{111} V\,.
\end{equation}
Hence, the magneto-asymmetry of $\mathcal{G}$, defined as
\begin{equation}\label{eq_asg}
\Phi_\mathcal{G} = \frac{1}{2} [\mathcal{G}(B)-\mathcal{G}(-B)]\,,
\end{equation}
depends only
on the asymmetry of the rectification coefficient $G_{111}$.
Moreover, we define the magneto-asymmetry of the screening potential:
\begin{equation}
\label{eq_asu}
\Phi_U = \frac{1}{2} [U(B)-U(-B)]\,.
\end{equation}
In turn, $\Phi_U$ depends on the symmetry properties of the
characteristic potentials. It is important to stress that injectivity 
$\overline{D}_{\alpha}(B)$ and the emissivity $\underline{D}_{\alpha}(-B)$ are evaluated in the equilibrium potential and therefore satisfy
the reciprocity relation 
$\overline{D}_{\alpha}(B) = \underline{D}_{\alpha}(-B)$.
The injectivities and the emissivities are in general
not even functions, in contrast to the total density of states 
(the sum of all injectivities or the sum of all emissivities).
As a consequence, the characteristic potentials 
(and thereby the potential landscape) which, as we will show, depend 
in an essential way on the injectivity are in general not even functions.
Therefore, the transport potential is in general not an even function 
of magnetic field.
Below, we prove that for a generic conductor (a chaotic cavity)
the fluctuations of the potential possess a magnetic-field asymmetry
and that, as a result, the fluctuations of the nonlinear
current--voltage characteristics are not
an even function of $B$.

\section{Nonlinear fluctuations of a quantum dot}
We consider a ballistic quantum dot in which in the classical limit the electron
trajectories are chaotic (see left inset of Fig.~\ref{dot})). This defines a chaotic cavity  which we connect to external reservoirs via two quantum point contacts  with adjustable
number $N_1$ and $N_2$ of propagating modes. Scattering in such
open cavities is successfully
described by random matrix theory~\cite{bee97,bar94}.

We assume that the cavity is in a perpendicular magnetic field 
with a total magnetic flux through the cavity of the order of one flux quantum,
$h/e$. This is sufficient to break time-reversal symmetry. The probability
distribution of the scattering matrix is uniform over the
unitary group (symmetry class $\beta=2$). 
On the ensemble average, the current is a linear function of the
applied voltage $V$. The average conductance is just the classical series 
resistance of the two quantum point contacts:
$\langle G \rangle = e^{2} N_1 N_2 /hN$ where $N=N_1 + N_2$. Weak
localization corrections vanish when $\beta=2$. 
Due to wave interference different members of the ensemble have a different 
{\em linear} conductance. For $N_1=N_2\gg 1$ the variance of the 
conductance fluctuations is 
\begin{equation}
\label{eq_fluclg}
{\rm var} G= \frac{e^4}{h^2} \frac{1}{8\beta} \,.
\end{equation}
Similar to the fluctuations in the linear conductance an individual 
cavity exhibits a fluctuating nonlinear $I$--$V$ characteristics.
Unlike the linear regime, which has been extensively studied~\cite{bee97},
the quantum fluctuations of the nonlinear conductance have found much 
less attention (see, however, Refs.~\onlinecite{alt86,web88,tab94,lin00,lof04}).
The energy scale of quantum interference effects is determined 
by the energy  
$h/\tau_d=N\delta$ where $\tau_d$ is the dwell time
for noninteracting electrons and 
$\delta$ the mean level spacing in the cavity. 
This energy is also the relevant energy scale 
for the nonlinear effects considered here. 

We next discuss in detail how we determine the transport potential 
$U$. A Poisson equation needs to be solved to obtain 
the characteristic potentials.
Here, we assume for simplicity that the potential can be described with
a single variable $U$ neglecting its spatial fluctuations within the cavity. 
In the magnetic field range considered here, 
the cavity is effectively zero-dimensional due to its isotropic 
scattering properties. We treat interactions within a random phase 
approximation which determines the Hartree potential with the help of a self-consistent 
effective interaction. \cite{but93} For the open cavity with ideal multi-channel
quantum point 
contacts such an approach can be justified rigorously. \cite{blf} In response to a voltage shift $V_\alpha$ in lead $\alpha$
a bare charge 
$Q^{bare}_\alpha=e^2 \overline{D}_\alpha V_\alpha$ is injected into the 
cavity. This excess charge generates a potential response which in turn 
leads to a screening charge
of {\em opposite} sign: $Q^{scr}=-e^2 D U$ determined by 
the total density of states $D = \overline{D}_1 + \overline{D}_2$.
In addition, the dot is coupled capacitively to a gate at voltage $V_g$
and geometric capacitance $C$. The excess charge on the cavity can now be 
expressed in two ways: consideration of the total charge response gives 
\begin{equation}\label{eq_cons}
Q=\sum_\alpha Q^{bare}_\alpha+Q^{scr} =
e^2 \overline{D}_1 V_1 + e^2 \overline{D}_2 V_2  - e^2 D U\,\,,
\end{equation}
and consideration of the Poisson equation gives 
\begin{equation}
Q= C (U- V_g ). 
\end{equation}
Therefore, using these two expressions to eliminate the total charge, 
we find for the potential
\begin{equation}
U=\frac{e^2 \overline{D}_1 V_1 + e^2 \overline{D}_2 V_2 + C V_g}{e^2 D+C}.
\end{equation}
Note that $U$ is the deviation from the equilibrium value $U_{eq}$.
Taking the derivatives of $U$ with respect to the voltage shifts 
$V_1 ,V_2 ,V_g$ gives the characteristic potentials $u_1 ,u_2, u_g$. 

We remark that different nonlinear $I$--$V$ characteristics are measured
depending on the way the cavity is biased. For the chaotic cavity 
considered here the characteristics differ just by a 
sample-to-sample fluctuation. 
Without loss of generality we consider   
$V_g = V_2=0$ and $V_1=V$. 
Thus, the expression for the differential conductance $\mathcal{G}$
effectively corresponds to the
two terminal case, Eq.~(\ref{eq_c2t}).

The approximation of a single potential used here neglects charge 
oscillations on the scale of the Fermi wave length. Therefore, we can 
use the WKB approximation to 
replace potential derivatives in Eq.~(\ref{gnonl})
with derivatives with respect to energy~\cite{chr96}.
Then,    
\begin{equation}\label{eq_g111}
G_{111}= - \frac{e^3}{h} \left.\frac{dT}{dE}\right|_{\rm eq}
(1-2 u_1)\,,
\end{equation}
where $T \equiv {\rm Tr}( s_{12} s_{12}^\dagger)$ 
is the transmission probability.
For the chaotic cavity the energy derivative of the transmission 
probability is random from one ensemble member to the other and for 
a sufficiently open contacts (large $N$-limit) is not 
correlated 
with the potential fluctuations. As a consequence, the ensemble average 
$\langle G_{111}\rangle=0$ vanishes. 
A magnetic-field asymmetry can develop only due to quantum fluctuations,
of the characteristic
potential $u_1$,
\begin{equation}\label{eq_u1}
u_1=\overline{D}_1\delta \frac{C_{\mu}}{C}\,.
\end{equation}
Here we have introduced the {\em electrochemical capacitance} \cite{but93} 
$1/C_{\mu} = 1/C + 1/e^{2}D$. For $e^{2}/C \gg \delta$ the 
fluctuations in the capacitance 
are very small~\cite{mello,bro97} and  
the actual density of states in the capacitance can be replaced 
by its ensemble average
$\langle D \rangle = 1/\delta $. Here $\delta$ is the level spacing 
in the cavity if the 
contacts to the reservoirs were closed.

From the characteristic potential we obtain the fluctuations
of the magneto-asymmetry of the screening potential, Eq.(\ref{eq_asu}), 
\begin{equation}\label{eq_fluctu}
{\rm var}\,\Phi_U=\frac{V^2\delta^2}{4} \left(\frac{C_{\mu}}{C}\right)^2
{\rm var} (\overline{D}_1-\underline{D}_1)\,,
\end{equation}
in terms of the fluctuations of the difference of the injectivity
and the emissivity.
Using the results of Ref.~\onlinecite{bro97} for the unscreened
emittance of a chaotic cavity in the large $N$ limit we
find
\begin{equation}\label{eq_fluctu2}
{\rm var}\,\Phi_U=\frac{N_1 N_2}{N^4}
\frac{V^2}{2} \left(\frac{C_{\mu}}{C}\right)^2\,.
\end{equation}
The size of the fluctuating magneto-asymmetry vanishes quickly
with increasing mode number (proportional to $1/N^2$ for a symmetrically
coupled cavity).  

We are now in a position to determine the magneto-asymmetry
of the fluctuating nonlinear conductance. First,
in the limit of a large number of modes the traces arising
in Eq.~(\ref{eq_g111}) can be decoupled, i.e.
we can disregard correlations between $dT/dE$ and $u_1$.
The {\em unscreened} nonlinear conductance
$- (e^3/h) dT/dE|_{\rm eq}$
changes sign randomly on the ensemble so that its average is zero~\cite{pol03}.
We find for the fluctuations ${\rm var} (dT/dE) = 8\pi^2 N_1^2 N_2^2/N^6$.
Using these results and Eqs.~(\ref{eq_g111}) and~(\ref{eq_fluctu2})
we find 
\begin{equation}
\label{eq_varphi}
{\rm var}\,\Phi_\mathcal{G}=\frac{16 e^6}{\hbar^2}
\frac{N_1^3 N_2^3}{N^{10}}
\left(\frac{V}{\delta }\right)^2
\left( \frac{C_{\mu}}{C} \right)^2\,.
\end{equation}
A characteristic feature is that ${\rm var}\,\Phi_\mathcal{G}$ is
maximal for perfect screening $C_{\mu} = C$, i.e., when
the charging energy of the dot is much larger than the 
mean level spacing, $e^2/C\gg \delta$. In the opposite limit of 
weak screening, the ratio $C_{\mu} /C$ tends to zero as $C$
tends to infinity.
Hence, the fluctuations
exhibit a magnetic field asymmetry only to the extent that
the potential fluctuations are an uneven function of $B$.
Importantly, the fluctuations of $\Phi_\mathcal{G}$
have an energy scale given by the applied voltage and
increase until $V$ is of the order of $h/\tau_d$. 
We remark that the fluctuations become smaller with 
increasing number of channels, suggesting that the effect
is observable in the quantum regime only (small number of modes).
This result has been confirmed in Ref.~\onlinecite{san04}
with a numerical simulation of the probability distribution  
of the rectification fluctuations (see Fig. \ref{dot}).
We have, thus, demonstrated that the fluctuations of the
differential conductance are not symmetric under reversal
of the applied magnetic field and that this is purely an 
interaction effect. 

Compared to the fluctuations of the linear conductance, Eq.~(\ref{eq_fluclg}),
the fluctuations of the differential conductance lack, notably,
a {\em universal} feature. Equation~(\ref{eq_varphi}) demonstrates
that the fluctuations of $\Phi_\mathcal{G}$ depend on
the microscopic details of the sample through the parameters
$\delta$ and $C$. How could one distinguish between both
types of fluctuations in a realistic experiment? We notice
that Eq.~(\ref{eq_fluclg}) shows a $\beta$ dependence
and no magnetic-field asymmetry for the linear conductance
while the nonlinear conductance magneto-asymmetry is zero
for $\beta=1$ ($B=0$) and maximal for $\beta=2$ ($B=h/eS$, with
$S$ the dot's area). In between, we expect a smooth crossover
from low to high magnetic fields. Moreover, the fluctuations in the 
linear conductance saturate in the asymptotic limit of the mode number
whereas the fluctuations of $\mathcal{G}$, as already emphasized, 
are vanishingly small for $N_1,N_2\gg 1$.

\section{Effect of dephasing}

For the chaotic cavity considered here nonlinearity arises only 
due to quantum interference. Therefore, it is interesting to investigate 
the role of phase breaking. Suppose that electrons 
retain their phase memory for a time $\tau_\phi$. Carriers dwell a time 
$\tau_d=h\delta/N$ inside the cavity. Note that the dwell time depends 
crucially on the size of the contacts $N = N_1 + N_2$. For $\tau_d \ll \tau_\phi$ we deal with a quantum coherent cavity described above, whereas 
for $\tau_d \gg \tau_\phi$ carriers are in the cavity long enough to loose
phase memory. 
 
Dephasing is induced by means of interaction with the
environment (phonons, radiation, impurities with internal dynamics,
fluctuations of gate voltages, etc.) or even with other carriers.
Here we simulate such processes
in a phenomenological way by introducing a fictitious voltage probe~\cite{but86b}
attached to the cavity~\cite{bar95,bro97b}. The current flowing through the probe
is zero, so charge conservation is maintained. Nevertheless,
when an electron enters the probe, it looses its phase memory
and the emerging electron is injected in the dot with an uncorrelated
phase~\cite{but86b}. 
The voltage probe is dissipative, carriers relax on average to the equilibrium distribution function at the voltage probe. Inelastic processes are not necessary 
for phase-breaking: phase breaking  can occur through quasi-elastic processes.
Here we will treat only the case of a dissipative voltage probe.

For simplicity, we assume full screening ($C\to 0$). With the voltage probe,
we have now  a {\em three}-lead cavity. Applying charge
conservation [see Eq.~(\ref{eq_cons})], we find the screening
potential
\begin{equation}\label{eq_uphi}
U=\frac{\overline{D}_1V_1+\overline{D}_2V_2+\overline{D}_\phi V_\phi}{D}\,,
\end{equation}
with $D=\overline{D}_1+\overline{D}_2+\overline{D}_\phi$ the total DOS.
Without loss of generality, we take $V_1=V$ and $V_2=0$.
The probe is connected via a contact with $N_\phi$ modes
and $V_\phi$ must be found by setting $I_\phi=0$:
\begin{equation}
V_\phi=\frac{G_{11}+G_{21}}{G_{11}+G_{12}+G_{21}+G_{22}} V\,,
\end{equation}
where $G_{\alpha\beta}$ are linear conductances of the three probe 
conductor. Upon
inserting this result in Eq.~(\ref{eq_uphi}) one can find
the fluctuations of $\Phi_U$ with
\begin{equation}
\Phi_U=\frac{\delta}{2}\left[ (\overline{D}_1-\underline{D}_1) V
+\overline{D}_\phi V_\phi(B) - \underline{D}_\phi V_\phi(-B)
\right] \,.
\end{equation}
In addition to the injectivity fluctuations as in Eq.~(\ref{eq_fluctu})
there are in ${\rm var}\,\Phi_U$ contributions due to
the probe's voltage fluctuations and the correlations
between the injectivity $D_\alpha$ and the probe's voltage $V_\phi$. The calculation is lengthy and
we just quote the final result for symmetric couplings
($N_1=N_2=N/2$):
\begin{equation}
{\rm var}\,\Phi_U= \left( \frac{V}{2} \right)^2
\frac{8N^3+8N_\phi^3+8N_\phi^2 N+ N_\phi N^2}{16 (N+N_\phi)^{5}} \,.
\end{equation}
For $N_\phi=0$ we recover our earlier expression,
Eq.~(\ref{eq_fluctu2}). When $N\gg N_\phi$ the leading-order
correction to Eq.~(\ref{eq_fluctu2}) is 
\begin{equation}
{\rm var}\,\Phi_U= \left( \frac{V}{2} \right)^2
\left[ \frac{1}{2N^2}-\frac{39}{16} \frac{N_\phi}{N^3}
\right] +\mathcal{O}(1/N)^4\,.
\end{equation}
As expected, weak dephasing leads to a reduction of the
observed magneto-asymmetry.  In the opposite limit for $N_\phi \gg N$
the magnetic field symmetry vanishes as $1/ N^{2}_{\phi}$.

The full calculation of the magneto-asymmetry of the differential 
conductance $\mathcal{G}$
is rather involved and we have presently only numerical 
results~\cite{san05}.
The results are in good agreement with the reduction
of ${\rm var}\,\Phi_\mathcal{G}$ with increasing coupling
to the fictitious probe.
Therefore, experiments carried out in cavities
attached to few-channel contacts and large
dephasing times are most promising for the observation
of the symmetry breaking discussed in this work.

\section{Conclusions}
In this work we have analyzed the magnetic-field asymmetries
arising in nonlinear mesoscopic transport which  signal a 
departure from 
Onsager's reciprocity relations. We have presented
a theory based on the scattering approach which predicts
that the magneto-asymmetry of the fluctuating rectification current
is caused exclusively
by the fact that, in general, the response of the
screening potential is not an even function of the magnetic
field. This is an interaction effect. We have investigated the size of the effect in
a generic mesoscopic conductor, an open quantum (chaotic) dot.
We predict fluctuations of the differential conductance
which are asymmetric with regard to magnetic field reversal.
We have discussed the nonuniversal form of the fluctuations
and its dependence on screening, applied voltage and energy.
Importantly, the magneto-asymmetry decreases rapidly with increasing
coupling to the reservoirs since for chaotic cavities the effect has a quantum
origin. Consequently, as we have shown here, the
fluctuations of the magneto-asymmetry are also suppressed with
increasing dephasing.

We have illustrated our theory with the help of a chaotic cavity. 
However, the main conclusions are completely general. 
Similar effects might be found in
molecular conduction junctions~\cite{mar93,che99,xue},
quantum wires~\cite{agr03} and carbon nanotubes~\cite{nit03}.
Our discussion shows that the magneto-asymmetry represents an 
important test of our understanding of nonlinear transport and 
its measurements can reveal information on the interaction strength 
in these structures.

\section*{Acknowledgments}
We thank H. Bouchiat, H. Linke, C. Marcus, M. Polianski, 
B. Spivak and D. Zumb\"uhl
for helpful discussions. We acknowledge support
from the Swiss National Science Foundation, the program MaNEP, 
the EU RTN under Contract No. HPRN-CT-2000-00144,
Nanoscale Dynamics, and the Spanish MECD through
the fellowship ``Ram\'on y Cajal''.

\section*{Note added in proof}
Two recent works present experimental data on magnetic field asymmetry
in field effect transistors \cite{riwy} and 
in carbon nanotubes \cite{cobd}. The data point to a classical 
effect different from the mesoscopic effects discussed here.

\end{document}